\documentclass[preprint]{aastex}

\usepackage{subfigure}
\usepackage{natbib}
\usepackage{graphicx}
\usepackage{multirow}
\usepackage{epstopdf}
\usepackage{amsmath}
\usepackage{url}
\usepackage{hyperref}
\usepackage{natbib}
\usepackage{lineno}
\usepackage{adjustbox}
\usepackage{tabularx}
\shorttitle{BAT GBM SGRB Comparison}

\begin{document}
	
\title{Do the \textit{Fermi} Gamma-Ray Burst Monitor and \\
		\textit{Swift} Burst Alert Telescope see the Same Short Gamma-Ray Bursts?}

\author{Eric~Burns\altaffilmark{*,1},
	Valerie~Connaughton\altaffilmark{2},
	Bin-Bin~Zhang\altaffilmark{3},
	Amy~Lien\altaffilmark{4},
	Michael~S.~Briggs\altaffilmark{1},
	Adam~Goldstein\altaffilmark{5},
	Veronique~Pelassa\altaffilmark{6}, and
	Eleonora~Troja\altaffilmark{7}		
	}

\altaffiltext{*}{\footnotesize Email: eb0016@uah.edu}
\altaffiltext{1}{\footnotesize University of Alabama in Huntsville, 320 Sparkman Drive, Huntsville, AL 35805, USA}
\altaffiltext{2}{\footnotesize Universities Space Research Association, Science and Technology Institute, 320 Sparkman Dr, Huntsville AL 35805}
\altaffiltext{3}{\footnotesize Center for Space Plasma and Aeronomic Research (CSPAR), University of Alabama in Huntsville, Huntsville, AL 35899, USA}
\altaffiltext{4}{\footnotesize NASA Postdoctoral Program Fellow, Goddard Space Flight Center, Greenbelt, MD 20771, USA}
\altaffiltext{5}{\footnotesize NASA Postdoctoral Program, Space Science Office, VP62, NASA/Marshall Space Flight Center, Huntsville, AL 35812, USA}
\altaffiltext{6}{\footnotesize Smithsonian Astrophysical Observatory, Po Box 97, Amado, AZ 85645}
\altaffiltext{7}{\footnotesize NASA Goddard Space Flight Center, Greenbelt, MD 20771}

\begin{abstract}
Compact binary system mergers are expected to generate gravitational radiation detectable by ground-based interferometers. A subset of these, the merger of a neutron star with another neutron star or a black hole, are also the most popular model for the production of short gamma-ray bursts (GRBs). The \textit{Swift} Burst Alert Telescope (BAT) and the \textit{Fermi} Gamma-ray Burst Monitor (GBM) trigger on short GRBs (SGRBs) at rates that reflect their relative sky exposures, with the BAT detecting 10 per year compared to about 45 for GBM. We examine the SGRB populations detected by \textit{Swift} BAT and \textit{Fermi} GBM. We find that the \textit{Swift} BAT triggers on weaker SGRBs than \textit{Fermi} GBM, providing they occur close to the center of the BAT field-of-view, and that the \textit{Fermi} GBM SGRB detection threshold remains flatter across its field-of-view. Overall, these effects combine to give the instruments the same average sensitivity, and account for the SGRBs that trigger one instrument but not the other. We do not find any evidence that the BAT and GBM are detecting significantly different populations of SGRBs. Both instruments can detect untriggered SGRBs using ground searches seeded with time and position. The detection of SGRBs below the on-board triggering sensitivities of \textit{Swift} BAT and \textit{Fermi} GBM increases the possibility of detecting and localizing the electromagnetic counterparts of gravitational wave events seen by the new generation of gravitational wave detectors.
\end{abstract}

\section{Introduction}
Gamma-Ray Bursts (GRBs) are traditionally classified by their duration using the $T_{90}$, which is the length of time between 5\% and 95\% of the cumulative fluence of the burst \citep{Kouv93}. The bimodal distribution of $T_{90}$ values suggests a separation of GRBs into two classes: short GRBs (SGRBs) with a $T_{90}$ $<$ $T_{90}^{th}$ and long GRBs (LGRBs) with a $T_{90}$ $>$ $T_{90}^{th}$, where $T_{90}^{th}$ is a separation threshold traditionally set to 2 seconds. Although some more physical classification methods (see e.g. \cite{Zhan07}, \cite{Zhan09}) and alternative measures of duration \citep{Zhan14} have been proposed, $T_{90}$ is still generally accepted as a good indicator of whether the GRB belongs to the long or short class, and possibly between two types of progenitors. LGRBs are widely believed to occur during the collapse of massive stars \citep{Woos06}. The detection of core-collapse supernova associated with the closest LGRBs provides strong observational confirmation of this theory. LGRBs are observed close to the center of star-forming galaxies, suggesting a young stellar population as their progenitors \citep{Fruc06}. 

Conversely, no supernova has been detected in association with a SGRB; \cite{Kann11} set stringent limits for a few nearby events. The properties of the host galaxies and the offset of the SGRBs far from the center of the host galaxies suggest an older population for the progenitors of SGRBs \citep{Troj10, Chur11, Berg14}. SGRBs are hypothesized to result from neutron star-neutron star (NS-NS) mergers or neutron star-black hole (NS-BH) mergers \citep{Ruff99, Ross03, Rezz11, Giac13}. This idea was recently strengthened by the possible detection of a kilonova (or mini-supernova; \citealt{Li98}, \citealt{Barn13}) following the short GRB 130603B \citep{Tanv13}.

Classification of individual GRBs as short or long can be complicated. The bimodal $T_{90}$ distributions of the two classes overlap and there is no clear boundary between the two populations; therefore, classification based on duration alone is less reliable than classification based on multiple observational criteria \citep{Zhan09}. \cite{Virg11} suggest the ensemble of short bursts result from a combination of merger events and another population of non-merger events. \cite{Horv09} postulate a third, intermediate class of bursts. Furthermore, there are energy and instrumental dependencies for both triggering efficiency and measurements of duration \citep{Qin13}. \cite{Band06} suggests that the image trigger algorithms used by \textit{Swift} BAT cause it to be less sensitive to SGRBs than a rate trigger, resulting in a decrease in ability to trigger on SGRBs. This is mitigated in practice by further analysis of rate triggers on the ground.

\begin{figure}
	\begin{center}
		\includegraphics[scale=0.5]{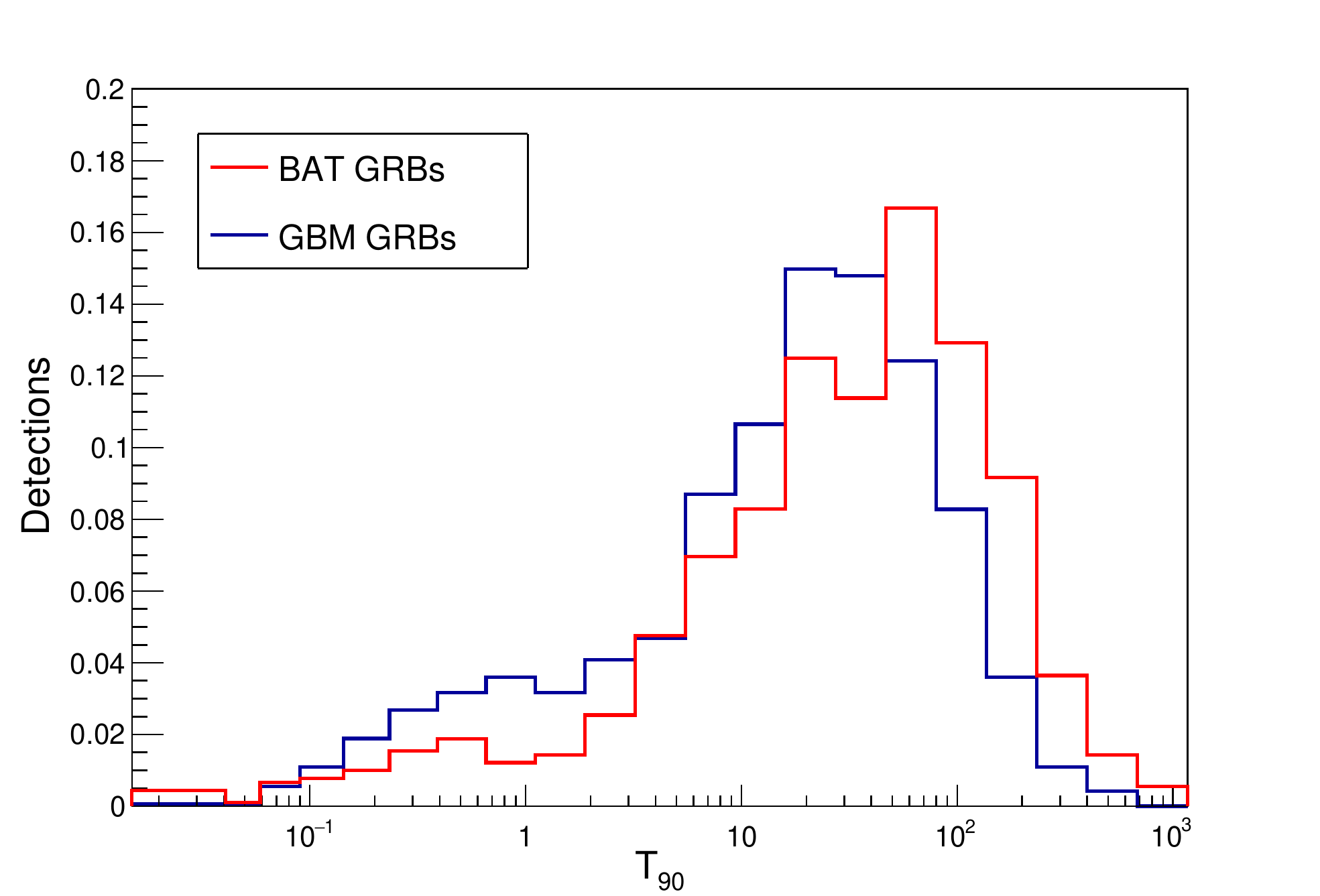}
	\end{center}
	\caption{$T_{90}$ distributions for \textit{Fermi} GBM and \textit{Swift} BAT. The distributions are normalized to unity.}
	\label{fig:botht90}
\end{figure}

The $T_{90}$ distributions for GRBs reported by the BAT and the GBM teams are shown in Figure \ref{fig:botht90}. The comparison of the respective ratios of LGRB detections to SGRB detections for each instrument suggests the two instruments are not seeing the same overall GRB population, as BAT's ratio of 10 to 1 is double GBM's ratio of 5 to 1. Determining how reliably instruments classify GRBs as distinct short and long populations that may have distinct progenitors helps uncover the differences between these populations and the properties of their gamma-ray emission. We study the SGRB populations detected by BAT and GBM to determine if the observed long-to-short ratios are a result of instrumental factors or if one instrument is seeing a population to which the other is less sensitive. Our SGRB samples for both instruments start when GBM began science operation on 14 July 2008 and end six and a half years later on 14 January 2015. Our SGRB sample for GBM is taken from the FERMIGBRST Catalog\footnote{\url{http://heasarc.gsfc.nasa.gov/W3Browse/fermi/fermigbrst.html}}, hosted on NASA's High Energy Astrophysics Science Archive Research Center (HEASARC), which contains all GBM triggered events classified as GRBs. Our GBM sample consists of 250 GRBs with GBM $T_{90}$ values less than 2 seconds. Our SGRB sample for BAT is built using the \textit{Swift} GRB Table and Lookup archive\footnote{\url{http://swift.gsfc.nasa.gov/archive/grb_table/}} and consists of 54 GRBs with listed BAT $T_{90}$ values less than 2 seconds. This includes those bursts found in ground analysis.

The detection of gravitational waves would be a breakthrough in physics. The association of SGRBs with mergers of compact objects has important consequences for the detection of electromagnetic radiation associated with gravitational wave (GW) radiation \citep{Bloo09, Pale13}. NS-NS mergers and NS-BH mergers are expected to emit gravitational waves during the inspiral phase of the collapse into a black hole \citep{Hugh09} in the frequencies detectable with the upcoming Advanced Laser Interferometer Gravitational-Wave Observatory (Advanced LIGO)/VIRGO detectors. A GW candidate with an associated SGRB would significantly strengthen the GW candidate as a true detection, in addition to confirming mergers as the source of SGRBs. If these merger events are the progenitors of SGRBs then we can estimate how many joint SGRB-GW candidates to expect between the Advanced LIGO/VIRGO experiments and the GRB detectors active in the Advanced LIGO/VIRGO era. Such estimates (e.g., \citealt{Siel13}) rely on the observed redshift distribution of SGRBs being an unbiased and uncontaminated sample in order to determine how many SGRBs occur within the expected detection horizon of Advanced LIGO/VIRGO. The number of SGRBs with known redshift is small, and nearly all of them are \textit{Swift} bursts. Contamination of this SGRB sample with non-merger events would corrupt this distribution and the associated predictions based on SGRBs with known redshift.

In Section 2 we discuss the instrumentation and triggering methodologies of both the \textit{Fermi} GBM and the \textit{Swift} BAT. In Section 3 we describe GBM SGRBs as viewed by the BAT, investigate if BAT could have observed more of these GRBs, and whether we can find evidence for these bursts in the BAT data. In Section 4 we describe the complementary search for BAT SGRBs in GBM. Section 5 investigates bursts with different duration classifications between instruments as well as detection differences for additional bursts which can be classified as short under different selection criteria. We discuss the broader implications of our findings in Section 6 and summarize our conclusions in Section 7.

\section{Instrumentation}
\subsection{The \textit{Fermi} Gamma-ray Burst Monitor}
The \textit{Fermi} GBM has twelve thallium-activated sodium iodide (NaI(Tl)) scintillation detectors that cover the energy range 8 keV to 1 MeV and two bismuth germanate (BGO) scintillation detectors that are sensitive between 200 keV and 40 MeV. The axes of the NaI detectors are oriented to optimize all-sky coverage and enable the localization of GRBs by comparing the relative observed source rates in each detector \citep{Meeg09}. The GBM can localize GRBs with an uncertainty of a few degrees \citep{Conn15}. 

The GBM has multiple triggering algorithms that act on varying energy ranges and timescales. For the 50-300 keV band the trigger algorithms operate on nine timescales from 16 ms to 4 s in steps of factor 2. The other three energy scales (25-50 keV, $>$100 keV, and $>$300 keV) currently operate on timescales starting at 16 ms with maximum timescales less than a second. The flight software triggers when it detects simultaneous increases in the count rates of two or more NaI detectors above an adjustable threshold specified in units of the standard deviation of the background rate. This information is detailed in \cite{Meeg09} and statistics about the GBM GRB triggers and the algorithms that detect them are provided in the GBM GRB catalogs, \cite{Paci12} and \cite{Kien13}.

\subsection{The \textit{Swift} Burst Alert Telescope}
The \textit{Swift} BAT is composed of a detector plane with 32,768 CZT detector elements, a coded-aperture mask above the detector plane, and a graded-Z fringe shield to reduce the background rate \citep{Bart05}. As compared to GBM, which has a large field of view with poor localization ability, a wider energy range, and spacecraft pointing dependent backgrounds, this design gives BAT lower backgrounds and the ability to detect weaker GRBs than \textit{Fermi} GBM. The BAT operates over the energy range 15-150 keV, with a 1.4 sr half-coded field of view (FoV), and can localize to an accuracy of a few arcminutes, which enables follow-up with narrow field instruments on-board \textit{Swift}. The BAT's FoV with localization capability is quantified by the partial coding fraction, which is the fraction of BAT detectors exposed to a position on the sky. In practice it has values between $\sim$0.0 and $\sim$1.0, where 1.0 is fully coded, 0.5 is half-coded, and 0.0 means the position is outside the region within which the BAT can localize an event.

The BAT trigger algorithms consider pre-burst and post-burst background intervals, the duration of the burst emission test interval (currently 4 ms to 16 s), the region of the detector plane that is illuminated (with different criteria for different combinations of detector quadrants), and the energy range (typically 4 different bandpasses). There are two stages to this trigger algorithm. The first stage is triggered by an increase in count rate above a set signal-to-noise ratio. After a rate trigger an image will be generated to confirm and localize the event. A second burst detection method is also run; the detector array count rate map is processed through the imaging algorithm on time intervals between 64 s and {\raise.17ex\hbox{$\scriptstyle\sim$}}5 minutes, and scanned for point sources. If the image signal-to-noise ratio for either algorithm exceeds a set threshold the event is confirmed as an unknown trigger source. If no source is found in the image after the rate trigger increase then it is considered a failed event. Ground inspection of failed rate triggers or data acquired when the spacecraft is slewing (when triggering is disabled) can result in further GRB detections. Details on the \textit{Swift} BAT are found in \cite{Bart05} and more details on the current BAT triggering algorithms can be found in \cite{Lien14}.

\section{The GBM SGRBs as viewed by \textit{Swift} BAT}
Within our sample time the GBM triggered on 250 GRBs that have GBM $T_{90}$ values less than 2 s, henceforth referred to as GBM SGRBs. Of these 250 GBM SGRBs the \textit{Swift} BAT detected 28, with the BAT triggering on-board for 22 of the 28. The 6 bursts found in ground analysis are GRBs 150101B, 141205A, 140402A, 120817B, 101129A, and 100216A. GRB 150101B was recovered from BAT slew data (Cummings et al., GCN 17267), GRB 101129A was a sub-threshold source on-board (Cummings et al., GCN 11436), while GRBs 141205A, 120817B, and 100216A were not found on-board (Cummings et al., GCNs 17137, 13692, and 10428, respectively). GRB 140402A was a marginal BAT source (Cummings et al., GCN 16071) despite being seen as a bright burst by GBM and occurring in a highly coded position. In contrast, the BAT found 219 of the 1524 GBM LGRBs in the same time period, a proportional detection rate increase of nearly 30\%.

\begin{figure}
	\begin{center}
		\includegraphics[scale=0.5]{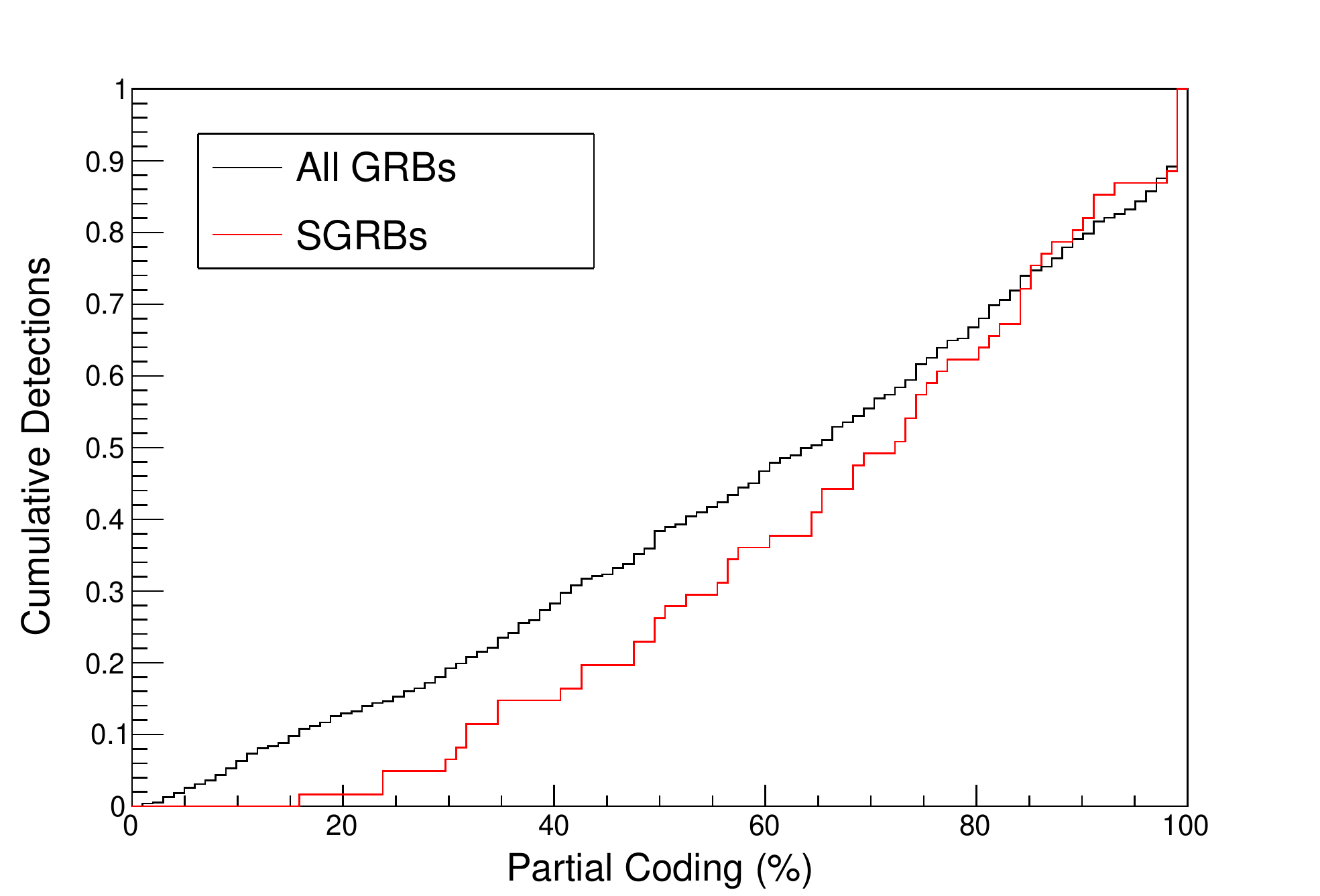}	\end{center}
	\caption{The cumulative fractional detection of triggered BAT GRBs as a function of partial coding values. It is evident that the BAT is more sensitive to LGRBs than SGRBs at lower partial coding values. The LGRB line closely tracks the line for all GRBs.}
	\label{fig:CPCD}
\end{figure}

In order to determine the minimum partial coding necessary to detect SGRBs we show in Figure {\ref{fig:CPCD}} the cumulative fraction of triggered BAT SGRBs as a function of partial coding, with partial coding values taken from GCN circulars. No BAT SGRBs found on board have been observed with a partial coding $<$ 10\%. The lowest partial coding value for a triggered SGRB is 16\% for the bright GRB 090510. The next lowest is 24\%. The ground detected GRB 120817B is the only BAT SGRB detected at a location with less than 10\% partial coding. It was an exceptionally bright SGRB that was well localized independent of \textit{Swift} by the Interplanetary Network. As neither of these conditions apply to the majority of GBM SGRBs, we impose a value of 10\% partial coding as a necessary minimum value for \textit{Swift} BAT to detect a SGRB.

We determined the exposure of BAT to GBM SGRBs by examining the region of the sky BAT was observing at GBM trigger time. An exact 10\% partial coding contour can be pulled from partial coding maps created from BAT survey or event data, the types of data BAT collects when outside of the South Atlantic Anomaly (SAA), not in safehold, and not slewing (though \textit{Swift} sometimes takes event data during slew). When the true partial coding maps are not available we use an approximated 10\% partial coding contour that is averaged from 14 real partial coding maps from our sample time. For these cases we use the continuous attitude files (Hans Krimm, private communication) to determine the pointing and roll of \textit{Swift} which allow us to map the approximate 10\% partial coding contour onto the sky. In the few cases where a GRB occurred in a time not covered by these attitude files we obtain the necessary information from the \textit{Swift} Observation Log\footnote{\url{http://www.swift.psu.edu/operations/obsSchedule.php}}, which is not as easily accessed in a machine readable format. In all cases (both real and approximated 10\% partial coding contours), we use the attitude files to obtain the operation flags at GBM trigger time to determine if the spacecraft was in SAA, safehold, or slewing. Over the course of \textit{Swift}'s mission the average number of active BAT detectors has decreased \citep{Taka14}. Despite this, the area contained within the 10\% partial coding contour is fairly constant over our sample time, with an average 1.96 sr FoV.

We find the maximum angular offset of the 10\% partial coding contour from the center of the BAT FoV is always less than 57 degrees. Our first step in checking the exposure of BAT to GBM SGRBs  was a rough cut based on the center of the GBM localization and the maximum angular offset from the BAT boresight to the 10\% partial coding contour. Since GBM can always locate bursts to one hemisphere of the sky the absolute maximum offset from the center of the GBM localization to the true location is 90 degrees. The first cut removed those GBM SGRBs whose GBM localization center is more than 147 degrees (the maximum 10\% partially coded position offset from the BAT boresight added to the maximum 90 degree localization error) from the BAT boresight at trigger time, the sum of these two maximum distances. This removed 18 GBM SGRBs. The second cut removed any GBM SGRB with a localization center further than five times the quadrature sum of the 1-sigma statistical error with a five degree systematic error to account for asymmetric localization contours; the five degree systematic is conservative as the average systematic uncertainty at 68\% confidence is 3.7 degrees \citep{Conn15}. This second cut removed a further 72 GBM SGRBs. Therefore, at least 90 GBM SGRBs from the total sample of 250 were not observable by BAT. 

The exposure of BAT to the remaining 160 GBM SGRBs was refined by calculating the fractional overlap of the GBM localization contours with the 10\% partial coding contours of BAT at the time of trigger. This sample of 160 includes the 28 GBM SGRBs that were also detected by BAT on-board or in ground analysis. The probability of a GBM SGRB's true location lying within the BAT 10\% partially coded FoV was calculated by summing the fractional overlap of the 1, 2, and 3 sigma contours multiplied by the probability of the true location being within each respective contour (68.27\%, 95.45\%, and 99.73\%, respectively) and adding the remainder (0.27\%) multiplied by the fraction of the 3 sigma contour within the 10\% partial coding contours. The results from this investigation are given in Table {\ref{table:GBMtoBATfull}}. The localization contours are a recent development by the GBM team and it was necessary to redo the localization to generate the contours. In some cases we found the new localization was significantly different from the original GBM localization, usually due to the original location being found with an earlier version of the localization algorithm. We consider the new locations to be more reliable and use them in this analysis. All GRBs in the GBM online catalogs will be updated with these new locations and contours.

We measure the uptime of \textit{Swift} BAT using the operation flags from the \textit{Swift} attitude files, counting all slews as downtime. BAT uptime during our six and a half year sample is 75.8\% of the total time; \textit{Swift} is in SAA 13.0\% of the time, slewing 12.9\% of the time, and in safe hold 0.1\% of the time (with some overlap among the three). With a 1.96 sr average field of view for partial coding values greater than 10\% and 75.8\% uptime, BAT has a total time averaged observable sky fraction of 11.8\%. The corresponding value for GBM is 59.8\%, obtained by examining the position history files over one spacecraft orbital precession period and removing times during SAA passage and times when a location is occulted by the Earth. Taking the 250 SGRBs detected by GBM and multiplying by the ratio of the total time averaged observable sky fraction (250 times 11.8\% over 59.8\%) implies an expected 49 SGRBs should be detected by the BAT above the GBM triggering threshold in our six and a half year sample. BAT actually detected, through a combination of on-board triggering and ground analyses, 54 SGRBs in this time, but as one was outside of the 10\% partially coded region (120817B), and one occurred during slew (150101B), we get an actual count of 52 BAT SGRBs after applying the same selection criteria. The consistency of the calculated and observed number of SGRBs implies the average sensitivity of the BAT to SGRBs out to 10\% partial coding is similar to the average sensitivity of GBM.

We expect 39 of the SGRBs detected by GBM to have occurred within BAT's FoV over our time sample (250 times the 1.96 sr field of view of BAT divided by the total area of the sky). This agrees with the 38 SGRBs expected from the sum of the areas of GBM localization contours within the 10\% partial coding, confirming we obtain an accurate approximation using our method. Removing the times that \textit{Swift} is in SAA, slewing, or in safehold, which is 24.2\% of the time, we expect BAT to have detected 29 or 30 GBM SGRBs. Of the 28 GBM SGRBs BAT did detect, one occurred during slew, so that the actual number of joint detections under these criteria is 27. Applying the same duty cycle criteria to the sum of the GBM localization overlap fractions gives 30 expected detections. Again, the predicted number of joint detections is similar to the observed number, implying that not only is the BAT detecting SGRBs at the same average sensitivity as GBM, but it is detecting the same population of SGRBs.

For the 57 GBM SGRBs that have non-zero GBM localization contour overlap with BAT's 10\% partial coding contours, and occurred during times that BAT could have collected event data (BAT uptime or slew time), but on which BAT did not trigger, we checked for evidence of the SGRBs in the BAT 64 ms lightcurves (David Palmer, private communication). BAT can observe bursts outside of its partially coded FoV but it cannot localize them; bursts in this region are not reported as detections by the BAT team. 14 of these 57 GBM SGRBs show no evidence of a burst in the BAT lightcurve. 13 of these 14 GBM SGRBs have more than a 90\% chance of having the true location outside of the BAT 10\% partially coded field of view, and have a significant chance of being outside BAT's entire field of view. The only exception is 140724533 which has an 87\% chance of occurring within the 10\% partially coded BAT FoV and only a small chance of occurring outside of the total BAT FoV. It is a very weak burst with very poor localization. The centroid of the localization is between two local minima, with the selected minimum strongly dependent on a user-selected source interval, the second being outside the BAT FoV. The other 43 GBM SGRBs are visible in the BAT lightcurves.

We examine the event data covering the GBM trigger time for 41 of the 43 GBM SGRBs visible in the BAT lightcurve but for which a detection was not reported. When \textit{Swift} BAT detects a rate trigger increase but fails to find a significant source in an uncatalogued location it collects either 3 or 10 seconds of event data (depending on whether the rate trigger was due to a slow or fast algorithm), referred to as failed event data. Most of these 41 GRBs occurred during failed event data, but a few occurred while BAT was collecting data for previous, unrelated triggers. If they occurred during failed event data and the failed event data starts close to the GBM trigger time then it is likely the rate increase was caused by the GRB. The results from a search of the event data for these 41 GBM SGRBs are given in Table {\ref{table:GBMtoBATevt}}.

We run the HEASOFT\footnote{\url{http://heasarc.nasa.gov/lheasoft/}} script batgrbproduct\footnote{\url{https://heasarc.gsfc.nasa.gov/ftools/caldb/help/batgrbproduct.html}} on these 41 GBM SGRBs. We first ran a blind search (no location seed), using trigger start and stop times (all input times for this script are in \textit{Swift}'s on-board MET) from visual inspection of the BAT lightcurves, and set the background start and stop times at 10 and 0.256 seconds before our input trigger start time. We visually inspected the lightcurve instead of adopting a more systematic method as the different binning in time and energy for the lightcurves of each instrument, the drift of \textit{Swift}'s on-board clock, and the non-negligible, variable light travel time between the two spacecraft all contribute to making automation difficult. We select the trigger start time as the first 64 ms bin clearly above background and the trigger stop time  as the return to background after the last peak evident in the BAT lightcurve, using the GBM lightcurve as a guide.  

GRB 140402A has the lowest SNR for a reported detection of a SGRB at 6.0, with a location consistent with the sub-degree location reported by the \textit{Fermi} Large Area Telescope (LAT). However, we note that the lowest SNR for a burst that was best localized by only GBM (which had the entire reported GBM localization within the BAT FoV) is 100216A with a ground analysis SNR of only 6.8 (Cummings et al., 10428). For those bursts with promising SNR regions, which we take as those positions with values above 5.5 that occur within the GBM localization contours, we further refined the analysis by varying the input times using the lightcurve as a guide. We set the minimum SNR for a possible location to be the true GRB at 6.0, but note this is not sufficient to claim a significant detection as we do not calculate the probability these locations are due to chance coincidence.

We find a few GBM SGRBs have positions with SNRs above 6.0 within the GBM localization contour, with the BAT lightcurves for that position consistent with the GBM lightcurves. The most promising signal of any GBM SGRB is for 090405663, with an SNR of 7.5, achieved by widening the source selection interval.  We do not calculate the significance of these detections but we note that  sub-threshold SGRBs in the BAT data likely exist, with localizations that are more accurate than GBM can achieve. Further investigation could refine the minimum SNR necessary to separate true signals from chance coincidence.

\section{The BAT SGRBs as viewed by \textit{Fermi} GBM}
Within our 6.5 year period, the \textit{Swift} BAT detected 54 GRBs which it classified as short under the standard $T_{90}$ threshold of 2.0 seconds, henceforth referred to as BAT SGRBs. Of these 54 BAT SGRBs \textit{Fermi} GBM triggered on 22. There are two reasons that prevent the GBM from observing a burst position at a specific time: occultation of the burst by the Earth and the passage of the \textit{Fermi} spacecraft through the SAA. We use the position history files accessible through the FERMIGDAYS catalog\footnote{\url{http://heasarc.gsfc.nasa.gov/W3Browse/fermi/fermigdays.html}} to determine whether a BAT SGRB occurred during \textit{Fermi} passage through the SAA or was occulted by the Earth at trigger time. Objects within 68 degrees of Earth center from \textit{Fermi}'s point of view are considered to be occulted by the Earth. This simplification ignores the variability in the height of the \textit{Fermi} spacecraft and the non-spherical shape of the Earth, but this does not affect our analysis as no BAT SGRB occurred between 67 and 69 degrees from the center of the Earth, which is a wider margin of error than necessary for these uncertainties. Of the 32 BAT SGRBs on which GBM did not trigger, 17 were occulted by the Earth and 9 occurred while GBM was in the SAA. As one burst was both Earth occulted and occurred during SAA transit, there are 7 BAT SGRBs that were observable by GBM but didn't trigger GBM. A summary of this search is given in Table {\ref{table:BatToGBMfull}}. 

The GBM does not have equal sensitivity across the unocculted sky. In order for the \textit{Fermi} LAT to achieve uniform exposure of the sky the satellite rocks the pointing between orbits. The GBM NaI detectors decrease in sensitivity as the angle from detector normal increases. The GBM team considers a given location adequately viewed by an NaI detector if it occurs above a detector response fraction of half of maximum, which occurs at 60 degrees \citep{Paci12}. The GBM NaI detector placements were designed to optimize the unocculted full-sky coverage given a rocking angle close to the zenith. The original rocking profile was 35 degrees north (relative to the local zenith) for one orbit, then 35 degrees south for one orbit. Due to spacecraft thermal considerations \textit{Fermi} gradually increased the rocking angle until settling on a rocking angle of 50 degrees. This has resulted in a portion of the sky, far from the LAT boresight, that has proportionally lower coverage by the GBM.

To search for sub-threshold signals in GBM data we use a new ground based search that is more sensitive than the on-board triggering algorithms. On-board triggering requires the count rates in the second brightest NaI detector to exceed a simple background model (average count rate over a moving 17 second time window) by $4.5 - 5.5 \sigma$ on one or more time-scales from 16 ms to 4.096s. The ground search finds weaker GRBs by using a non-parametric fitting algorithm to model the background over longer timescales and uses a lower signal threshold. The ground search identifies GRBs that have significant signal in only one detector and those in which multiple detectors view the GRB but the significance of the second brightest detector does not reach the on-board triggering requirements. In each of these cases, false positives are eliminated by requiring that the relative rates in the 12 NaI detectors be compatible with an astrophysical origin. This test involves feeding the background-subtracted rates into our source localization software and applying a $\chi^2$ cut to the best fit, as described in \cite{Conn15}. We estimate that the ground search can find GRBs  1/3 as bright as the faintest triggered SGRB, resulting in a doubling of the SGRBs detected by GBM. Candidate vetting and automated false-positive identification are in progress and the search method and results will be presented in Zhang et al. (in prep).

We used this algorithm to search for evidence of these 7 BAT SGRBs, those that were observable but did not trigger GBM, in the GBM time-tagged-event (TTE) data, or continuous time (CTIME) data (see \citealt{Meeg09} for more information on GBM data types).  BAT SGRBs more recent than November 2012 were processed using continuous TTE data, while for older bursts we were limited to daily CTIME data (as continuous TTE data is not available prior to November 2012). We use TTE data when available because its finer time resolution is better matched to the shortest GRBs than the native 0.256 s binning of CTIME.

We find 4 of these 7 BAT SGRBs are detected and 3 are not detected, as shown in Table {\ref{table:BatToGBMFS}} which places the BAT SGRBs in context with the overall SGRB population detected by the BAT. The brightness and hardness percentiles are derived from the sample of all \textit{Swift} BAT SGRBs. Every BAT SGRB that didn't trigger GBM occurred in a position in the BAT FoV that was at least 50\% coded, where BAT is most sensitive to weaker events. Table \ref{table:BatToGBMFS} also includes the offset angle from the LAT boresight, the number of GBM NaI detectors that were viewing the burst location, and the two lowest observation angles from NaI detector normal. We exclude all detectors whose view was obstructed by the LAT or the spacecraft. 

In order to explore the possibility that these 7 BAT SGRBs represent a special class of BAT SGRBs we qualitatively investigate their duration, fluence, flux, power law index, and partial coding values as compared to the rest of the BAT SGRB sample. The $T_{90}$ values are neither systematically short or long, the fluence percentiles are fairly distributed, the power law indices are among the lowest and highest of the overall BAT SGRB power law index distributions, and the partial coding values are all above half, but this is not extremely high when compared to the total partial coding distribution for BAT SGRBs. The only observational quantity that appears to be systematically different from the full distribution is the peak flux, as only one of the seven values is above the total sample median. With a larger sample we might draw some statistically significant conclusions about these properties other than saying qualitatively that the SGRBs undetected by GBM appear to be weaker but otherwise similar to the overall BAT SGRB sample.

Looking at these 7 BAT SGRBs individually: GRB 140606A didn't trigger GBM due to low peak flux. GRB 120403A is a SGRB of average brightness but didn't trigger GBM. The lightcurve shows a relatively flat profile; GBM may not have triggered due to the lack of an impulsive spike given that GBM's sensitivity to SGRBs is derived from its short triggering windows. GRB 140516A is exceptionally weak, with both brightness measures in the bottom tenth for BAT SGRBs. These three bursts were sub-threshold for an on-board trigger but were found in ground based searches due to favorable viewing geometry. GRB 090305A was a high peak flux GRB that was also found in ground analysis despite being well viewed in only one detector. That lack of an on-board GBM trigger is explained by triggering requiring two detectors significantly above background. GRBs 110112A and 090815C were both favorably viewed by only one NaI detector but were not found in the ground analysis due to being weak bursts. They might be recoverable with finer time resolution data, but we lack the TTE data at the times these bursts occurred. GRB 140129B has an average fluence, modest brightness, and favorable geometry to GBM but the ground based search finds no signal. We can clearly see the burst in the GBM daily data, but only below 50 keV. As the ground based search is optimized for the 50-300 keV range we would not expect it to find this burst.

\section{Different Duration Classifications}
In addition to lack of detection, disagreement in classifications of bursts into the short or long class can also bias the respective SGRB populations. Again, we cannot calculate absolute misclassification, only the relative disagreements in classification between instruments. Due to the statistical uncertainties in the measurement of $T_{90}$ we would expect some disagreement.

Four of the GBM SGRBs that also triggered the BAT were classified as long by BAT. These are GRBs 090927 and 131128A (with BAT $T_{90}$ values of 2.2 $\pm$ 0.4 and 3.0 $\pm$ 1.41, respectively), which are consistent with a short classification within the duration measurement error, and GRBs 090531B and 140209A, which are suggested to belong to the class of SGRBs with extended emission as described in \citealt{Gehr07} (with BAT $T_{90}$ values of 80 $\pm$ 23 and 21.3 $\pm$ 0.8, respectively). The only BAT SGRB classified as long  by GBM (140320A) has a GBM $T_{90}$ of 2.30 $\pm$ 1.52 s, which is consistent with a short classification within errors. Therefore, the disagreement in short or long classification between instruments is likely to be due to statistical chance and is within $T_{90}$ error except when the burst is thought to be a SGRB with extended emission.

For completeness we conducted additional searches for bursts that may be believed to belong to the short population under different selection criteria. The first additional search was for bursts thought to be SGRBs with extended emission. Our sample consisted of GRBs 090531B, 090915, 090916, 110402A, and 111121A. GBM triggered on, and classified as short, GRBs 090531B and 140209A. Evidence for the extended emission for these two bursts is not found in GBM data due to the soft, weak extended emission being indistinguishable from background. GBM also triggered on GRBs 090915 and 110402A which have GBM $T_{90}$ values longer than 2 s, suggesting GBM also found the extended emission which was included in the $T_{90}$ interval. GBM didn't trigger on GRBs 090916 and 111121A due to either SAA passage or occultation by the Earth.

The second additional search was for those GRBs with a BAT $T_{90}$ below 3 seconds, as has sometimes been used as the $T_{90}^{th}$ \citep{Gehr07}. This provides 13 additional bursts of which 5 triggered GBM on-board and 4 were unobservable due to Earth occultation or SAA transit. Of the 5 that triggered both instruments one is classified by GBM as a SGRB, 3 of the others have $T_{90}$ values with error bars that overlap the 2.0 s threshold, and the last is classified as a long GRB with a GBM $T_{90}$ of 2.624 $\pm$ 0.326 s. Four bursts were observable by GBM but did not trigger GBM. Three were subsequently found in a ground analysis using CTIME data (two with strong signals, the other with a weak signal). The remaining burst occurred near the start of a bright solar flare and was not found in CTTE data.

The last additional search was for bursts that have GCN reported $T_{90}$ values under 2 seconds but do not have $T_{90}$ values listed in the \textit{Swift} GRB Table and Lookup Archive. This can be due to bursts being found in ground analysis for which BAT cannot find a reliable $T_{90}$, or bursts with observation limitations. GRBs 100216A, 120817B, and 130716A triggered GBM and have GBM $T_{90}$ values less than 2 seconds. GRBs 100224A and 130822A did not trigger GBM but both occurred during either SAA passage or were occulted by the Earth. Therefore, our results for the standard SGRB duration selection criteria also apply to the bursts covered by these three alternate selection criteria.

\section{Discussion}
\textit{Swift} BAT detected 22 GBM SGRBs on-board and 6 on the ground. In this analysis we identify a few additional GBM SGRBs that were possibly observed by BAT. The \textit{Swift} BAT team's ground search is therefore capturing the majority of SGRBs that fail to trigger BAT, down to GBM's sensitivity. There does appear to be a deficit of SGRBs detected by BAT as compared to LGRBs at partial coding values below 80\% but this deficit arises mostly from BAT missing SGRBs below the triggering threshold of GBM. The large difference between BAT and GBM in the ratios of LGRB and SGRB detections is due to the BAT's relative increased sensitivity to LGRBs. 

\textit{Swift} BAT detects weaker short bursts than GBM, when they occur in BAT's highly partially coded region. The average BAT sensitivity out to 10\% partial coding is close to the average GBM sensitivity for SGRBs so that the ratio of GBM to BAT total SGRB detections directly reflects their sky coverage ratio.

As there are 22 BAT SGRBs that triggered both instruments and 7 observable BAT SGRBs that did not trigger GBM, GBM's on-board trigger algorithm misses around 1 in 4 of the SGRBs detectable by BAT. We find evidence in GBM data for four of these bursts in a more sensitive ground analysis. Therefore, our ground based analysis can currently recover more than half of the BAT SGRBs on which GBM doesn't trigger in a seeded search and the GBM cannot find evidence for fewer than 10\% of the total BAT SGRB population. This non-detection rate is even lower now with the availability of continuous TTE for the ground search. Seeding the search allows for the detection of more sub-threshold SGRBs, as those bursts with weak signals in Table {\ref{table:BatToGBMFS}} were not strong enough in GBM to claim independent discovery. As these all occurred above 50\% partial coding, both instruments are missing some weak short bursts, the BAT in its region of low partial coding and the GBM for weaker bursts across its FoV, particularly those viewed with favorable geometry by only one detector.

\cite{Brom13} hypothesize that BAT's sensitivity to low fluence GRBs has two effects on the BAT $T_{90}$ distribution: (i) the BAT sees many weak LGRBs that are not detectable by GBM, making the short-to-long ratio lower for BAT than in GBM and (ii) the low fluence LGRB population contaminates the SGRB population, meaning the long end of the SGRB population has significant numbers of non-merger events that are the short end of the LGRB population of collapsars. The paper defines the contamination rate as the percent of bursts below a $T_{90}^{th}$ that are actually due to collapsars. Using the standard $T_{90}^{th}$ of 2 s they give the contamination rate of the GBM SGRB sample as around 15\% and the contamination rate of the BAT SGRB sample as about 40\%. They suggest a $T_{90}^{th}$ of about 0.8 seconds for the BAT sample will reduce the contamination and yield an SGRB sample that has a similar fraction of merger (or non-collapsar) events to that detected by GBM. We find that of the 7 SGRBs detected by BAT that did not trigger GBM, 5 have a combination of poor geometry to GBM and low peak flux. These 5 all have BAT $T_{90}$ values less than their suggested $T_{90}^{th}$ of 0.8 seconds. This leaves only two bursts that would be excluded by the 0.8 s $T_{90}$ cut and that might be considered contamination of the BAT SGRB sample above the contamination levels of the GBM SGRB population. That 2 of the 29 SGRBs detected by BAT might be non-merger events is not compatible with the hypothesis of \cite{Brom13} that 40\% of the SGRB BAT sample is non-merger contamination compared to only 15\% of the GBM sample; therefore, imposing the 0.8 s cut on $T_{90}$ will remove many genuine merger candidates without significant purification of the sample of merger events. This has important implications when considering the redshift distribution of SGRBs.

\begin{figure}
	\begin{center}
		\includegraphics[scale=0.7]{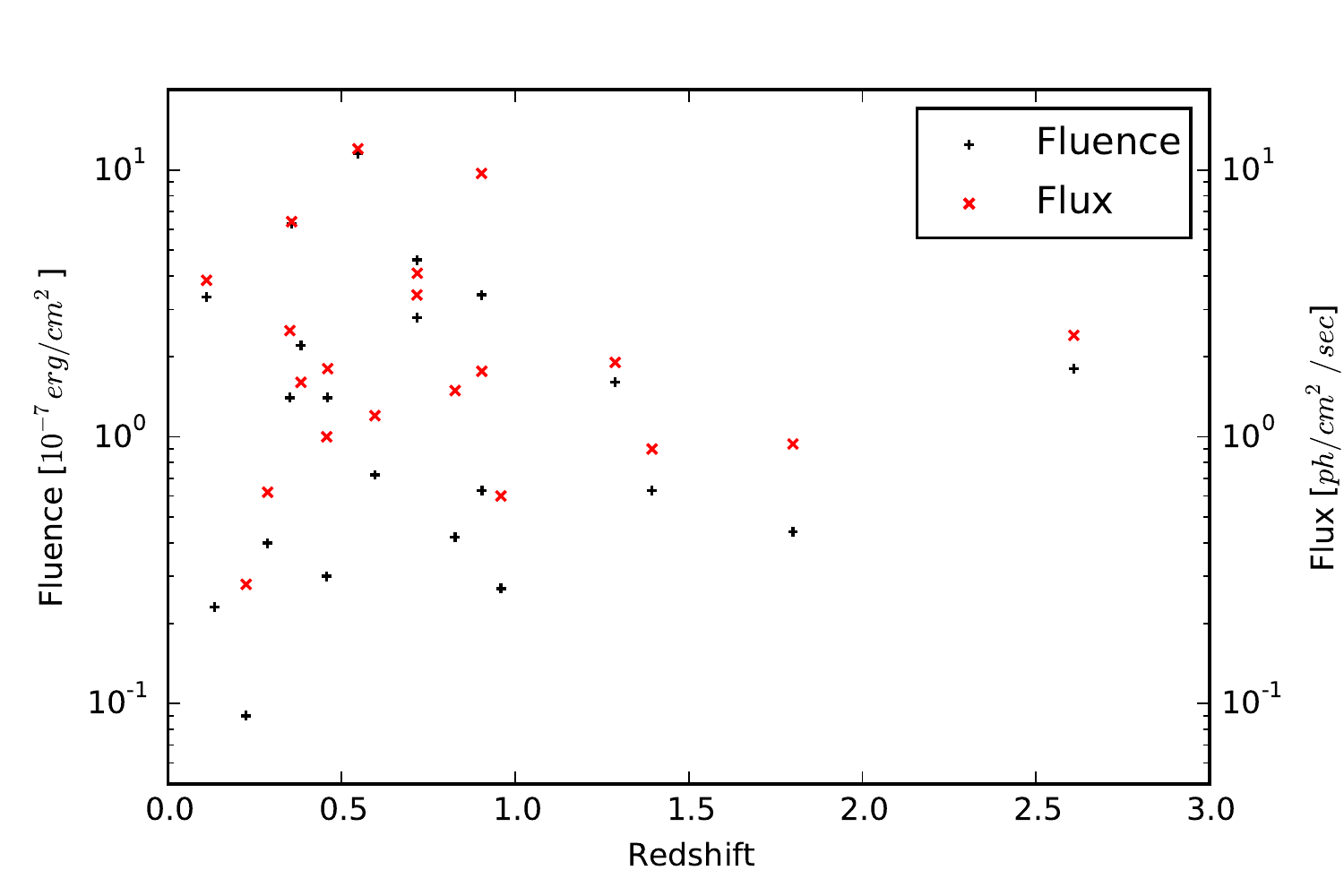}
	\end{center}
	\caption{Brightness measures for \textit{Swift} BAT bursts with measured redshift. Both measures of brightness are in the 15-150 keV range.}
	\label{fig:zbright}
\end{figure}

Figure {\ref{fig:zbright}} shows there is no redshift dependence on brightness of SGRBs, especially for weak bursts; therefore, the seven BAT SGRBs that were not detected by GBM were not particularly far away. The redshift distribution from BAT SGRBs is likely to be as good a representation of the underlying GBM SGRB redshift distribution as it is for the underlying BAT SGRB redshift distribution. Additionally, we note that extremely close events are not necessarily bright. The closest SGRB with a confirmed redshift is 150101B which has a fluence among the weakest of BAT SGRBs with redshift measurements. In fact, only 8.5\% of all BAT SGRBs have lower fluence than 150101B. In GCN 17267 it is suggested that GRB 150101B is an unusually soft event in BAT data. We do not find this to be the case when looking at the GBM data. The event is too weak for a constraining spectral fit but the count rates are peaked in the 100-300 keV range typical for SGRBs (many LGRBs are peaked in the 50-100 keV range). GRB 150101B was detected by BAT while BAT was slewing and it is possible that this, combined with the unusually weak fluence, led to the disagreement in spectral results between BAT and GBM. Therefore, the ground based searches for both instruments that pick up sub-threshold events may discover untriggered SGRBs within the Advanced LIGO/VIRGO horizon that would otherwise go undetected and unassociated with GW candidate discoveries.

\section{Conclusion}
We conclude that the \textit{Swift} BAT and the \textit{Fermi} GBM are observing the same SGRB populations. There are some missed detections but these are almost entirely due to those GRBs occurring in poorly viewed regions for the instrument that missed them and well viewed regions for the instrument that detected them. We have shown that \textit{Swift} BAT, through a combination of on-board algorithms and ground-based analysis, detects nearly all SGRBs above the \textit{Fermi} GBM trigger threshold within its 10\% partially coded FoV. BAT detects SGRBs below GBM's triggering threshold in its fully-coded FoV, suggesting more bursts could be recovered from locations at lower partial coding values. We find that GBM's new ground-based analysis is capable of recovering a large fraction of the SGRBs that failed to trigger GBM down to BAT's optimal sensitivity, particularly with the availability of CTTE since November 2012. The 4 SGRBs found in GBM data for the first time as a result of this analysis show the promise of this new technique for finding additional SGRBs. We also find that while BAT is detecting nearly all SGRBs down to GBM's sensitivity it may be possible for them to find evidence for more SGRBs in their data, for example 090405663. We find that both instruments can detect more SGRBs when seeded with the time and location of a detection from another instrument. This is particularly important when considering searches of GBM and BAT data for events in coincidence with Advanced LIGO/VIRGO gravitational wave candidates.

The nature of SGRB progenitors is still uncertain, though compact binary mergers are becoming more convincing with the association of SGRBs with older stellar populations and the recent possible detections of kilonovae associated with GRBs. We have shown that the SGRB populations detected by \textit{Swift} BAT and \textit{Fermi} GBM are similar. The redshift distribution of \textit{Swift} SGRBs is better characterized, but \textit{Fermi} GBM detects nearly five times as many SGRBs, making a joint EM-GW detection more likely to occur with GBM. Our analysis shows that we can confidently use the redshift distribution of the SGRB population detected by \textit{Swift} to calculate the expected numbers of SGRBs detected by both \textit{Swift} and by GBM within the Advanced LIGO/VIRGO horizon.

\acknowledgments
We would like to acknowledge the contributions of two people. David Palmer who generated BAT lightcurves, allowing us to further investigate GBM SGRBs as viewed by the BAT, and Hans Krimm who compiled continuous attitude files for \textit{Swift}, saving us a great deal of time. We additionally recognize the efforts of the HEASARC in providing the searchable databases that were the source of our data. The GBM members acknowledge support from GBM through NNM11AA01A/MSFC. Eric Burns acknowledges support through NASA Swift GI grant NNX15AC05G.

\renewcommand{\arraystretch}{0.75}
\begin{table}
	\scriptsize
	\centering`
	\begin{adjustbox}{width=\textwidth,totalheight=\textheight,keepaspectratio}
		\begin{tabular}{crc|crc|crc|crc}
			\hline
			GRB Trig. &   Area    &   Op.     & GRB Trig.   &   Area    &   Op.     & GRB Trig. &   Area    &   Op.     & GRB Trig. &   Area    &   Op.     \\
			or Name	  &	  (\%)    &	Flag	& or Name	&	(\%)	&	Flag	& or Name	&	(\%)	&	Flag	& or Name	&	(\%)	&	Flag	\\
			\hline
			150101B	&	93.9	&	Slew	&	130706900	&	0.0	&		&	111207512	&	0.0	&		&	100117A	&	99.9	&		\\
			150101A	&	56.6	&		&	130705398	&	0.3	&	SAA	&	111117A	&	99.8	&		&	091224373	&	0.0	&		\\
			141230871	&	0.0	&	Slew	&	130626452	&	86.2	&		&	111103948	&	16.3	&		&	091223191	&	7.6	&		\\
			141208632	&	0.2	&	SAA	&	130626A	&	0.0	&		&	111024896	&	0.0	&	SAA	&	091122163	&	80.4	&		\\
			141205A	&	91.9	&		&	130515A	&	100.0	&		&	111022854	&	0.0	&		&	091012783	&	0.0	&		\\
			141202470	&	22.9	&	Slew	&	130504314	&	0.7	&		&	111001804	&	0.5	&		&	091006360	&	0.0	&	SAA	\\
			141126233	&	55.3	&		&	130503214	&	9.2	&		&	110916016	&	2.4	&	Slew	&	090927	&	88.4	&		\\
			141113346	&	12.2	&	Slew	&	130416770	&	0.0	&		&	110728056	&	0.0	&	SAA	&	090909854	&	0.0	&		\\
			141111435	&	0.0	&	SAA	&	130404877	&	0.0	&		&	110705151	&	85.0	&	Slew	&	090819607	&	1.2	&		\\
			141105406	&	100.0	&	Slew	&	130307126	&	0.0	&		&	110605780	&	0.0	&		&	090717111	&	3.5	&		\\
			141102112	&	0.5	&		&	130219626	&	5.4	&		&	110529034	&	0.2	&	SAA	&	090621B	&	99.5	&		\\
			141031998	&	0.0	&		&	130127743	&	1.1	&	Slew	&	110509475	&	26.0	&	Slew	&	090620901	&	0.0	&		\\
			141020439	&	0.0	&		&	121127914	&	10.4	&	SAA	&	110422029	&	41.9	&	SAA	&	090616157	&	0.0	&		\\
			141011282	&	0.0	&	Slew	&	121124606	&	4.1	&	Slew	&	110420B	&	93.0	&		&	090531B	&	100.0	&		\\
			140831215	&	31.6	&		&	121112806	&	0.0	&		&	110409179	&	0.0	&		&	090510	&	41.6	&		\\
			140807500	&	0.0	&		&	121004211	&	0.0	&	SAA	&	110213876	&	4.1	&		&	090429753	&	100.0	&	Slew	\\
			140724533	&	87.2	&		&	120916085	&	0.0	&	Slew	&	110131780	&	0.0	&	SAA	&	090427644	&	0.0	&		\\
			140720158	&	0.0	&		&	120831901	&	0.1	&	SAA	&	101224A	&	80.5	&		&	090412061	&	0.0	&		\\
			140710537	&	0.0	&		&	120822628	&	91.7	&		&	101216721	&	0.0	&		&	090405663	&	16.6	&		\\
			140624423	&	99.2	&	SAA	&	120817B	&	12.2	&		&	101208203	&	0.0	&		&	090308734	&	0.1	&		\\
			140619490	&	0.0	&	SAA	&	120805706	&	0.0	&		&	101204343	&	62.0	&		&	090206620	&	0.0	&		\\
			140616165	&	12.8	&	Slew	&	120727354	&	0.0	&		&	101129726	&	0.7	&		&	090126245	&	0.0	&		\\
			140610487	&	4.9	&		&	120701654	&	18.3	&		&	101129A	&	35.9	&		&	081230871	&	0.0	&		\\
			140605377	&	5.5	&		&	120616630	&	0.0	&	SAA	&	101119685	&	1.6	&		&	081229187	&	0.0	&		\\
			140526571	&	20.3	&		&	120612687	&	0.0	&	SAA	&	101031625	&	0.1	&	SAA	&	081226509	&	100.0	&	Slew	\\
			140518709	&	0.0	&		&	120608489	&	0.4	&		&	100811108	&	0.0	&		&	081226A	&	70.8	&		\\
			140511095	&	15.2	&		&	120524134	&	0.0	&	SAA	&	100805300	&	2.3	&		&	081216531	&	0.0	&		\\
			140501139	&	46.3	&		&	120509619	&	97.5	&		&	100722291	&	0.0	&		&	081213173	&	0.6	&		\\
			140402A	&	100.0	&		&	120429003	&	9.0	&		&	100719311	&	0.0	&		&	081209981	&	0.0	&		\\
			140209A	&	100.0	&		&	120415891	&	0.0	&	SAA	&	100629801	&	0.0	&		&	081204517	&	6.3	&		\\
			140129499	&	5.4	&		&	120410585	&	0.0	&		&	100625A	&	100.0	&		&	081115891	&	0.0	&		\\
			140110411	&	0.0	&		&	120327418	&	0.0	&		&	100616773	&	0.0	&		&	081113230	&	59.1	&		\\
			140109771	&	13.5	&		&	120314412	&	0.0	&		&	100516396	&	20.7	&	SAA	&	081102365	&	0.0	&		\\
			131217108	&	0.0	&	Slew	&	120302722	&	0.0	&		&	100411516	&	0.0	&		&	081101	&	22.8	&		\\
			131128A	&	89.3	&		&	120222021	&	0.6	&	Slew	&	100301068	&	0.0	&	SAA	&	081024891	&	0.2	&	Slew	\\
			131006367	&	86.1	&		&	120212353	&	0.0	&		&	100223110	&	3.0	&		&	081024A	&	97.3	&		\\
			131004A	&	98.3	&		&	120205285	&	0.0	&		&	100216A	&	99.9	&		&	080905A	&	98.8	&		\\
			130912A	&	24.4	&		&	120129312	&	0.6	&		&	100208386	&	5.2	&		&	080802386	&	0.0	&	SAA	\\
			130802730	&	4.6	&	SAA	&	120101354	&	0.9	&		&	100206A	&	50.4	&		&	080725541	&	14.5	&		\\
			130716A	&	99.3	&		&	111222619	&	0.0	&		&	100204858	&	0.0	&		&	080723913	&	0.0	&	SAA	\\
			
			\hline			
		\end{tabular}
	\end{adjustbox}
	\caption{GBM SGRBs as viewed by BAT. A GRB name is given only for bursts which were also seen by BAT. The area is the percent of the GBM localization contours with the 10\% BAT coded FoV. The flags are the operating flags of BAT.}
	\label{table:GBMtoBATfull}
\end{table}

\begin{table}
	\centering
	\scriptsize
	\begin{tabular}{c c c c c}
		\hline
		
		GRB	&	Flag   & Event Data Start [s]	&	Event Data Total [s]	&	SNR max	\\
		\hline
		141202470	&	Slew	&		&		&		\\
		141126233	&		&	-1.84	&	10.08	&	4.8	\\
		141113346	&	Slew	&	-54.78	&	183.33	&	5.5	\\
		141105406	&	Slew	&		&		&		\\
		140616165	&	Slew	&		&		&		\\
		140605377	&		&	-2.1	&	10.08	&	4.1	\\
		140526571	&		&		&		&		\\
		140511095	&		&	0	&	3.13	&	5.6	\\
		140501139	&		&	0.05	&	3.13	&	5.5	\\
		140129499	&		&		&		&		\\
		140109771	&		&	-2.5	&	10.06	&	6.0	\\
		131006367	&		&		&		&		\\
		130504314	&		&	-2.51	&	10.08	&	4.6	\\
		130416770	&		&		&		&		\\
		130219626	&		&	-1.24	&	10.04	&	5.4	\\
		130127743	&	Slew	&		&		&		\\
		120822628	&		&		&		&		\\
		120701654	&		&		&		&		\\
		120608489	&		&	-1.3	&	10.07	&	5.1	\\
		120509619	&		&	-2.19	&	10.1	&	5.2	\\
		120222021	&	Slew	&		&		&		\\
		120129312	&		&		&		&		\\
		120101354	&		&	-1.74	&	10.08	&	4.4	\\
		111001804	&		&		&		&		\\
		110705151	&	Slew	&		&		&		\\
		110509475	&	Slew	&	-5.94	&	120.08	&	5.8	\\
		110213876	&		&	-2.66	&	10.06	&	5	\\
		101204343	&		&	0.57	&	3.12	&	5.5	\\
		101129726	&		&		&		&		\\
		100805300	&		&		&		&		\\
		100223110	&		&		&		&		\\
		091223191	&		&	-2.37	&	10.12	&	6.3	\\
		091122163	&		&	-2.28	&	10.02	&	5.3	\\
		090429753	&	Slew	&		&		&		\\
		090405663	&		&	-2.63	&	10.1	&	7.5	\\
		090308734	&		&		&		&		\\
		081226509	&	Slew	&		&		&		\\
		081213173	&		&	-0.33	&	3.06	&	4.3	\\
		081204517	&		&	0.69	&	3.06	&	5	\\
		081113230	&		&	-1.25	&	10.05	&	5.1	\\
		080725541	&		&	-1.97	&	10.08	&	6.1	\\
	\end{tabular}
	\caption{Event information for GBM SGRBs untriggered in BAT. We list the GRB, the event start time relative to GBM trigger, and the total number of seconds during this event file. We also report the highest SNR within the GBM contours. Event information is not given for cases where the processing script didn't complete and report an SNR $>$ 4.0.}
	\label{table:GBMtoBATevt}
\end{table}

\renewcommand{\arraystretch}{0.75}
\begin{table}
	\small
	\begin{adjustbox}{width=\textwidth,totalheight=\textheight,keepaspectratio}
		\begin{tabular}{c c c c c | c c c c c}
			\hline
			SGRB	&	GBM Trig.	&	E.O.	&	S.A.A.	&	Missed	&	SGRB	&	GBM Trig.	&	E.O.	&	S.A.A.	&	Missed	\\
			\hline
			150101B 	&	150101641 	&		&		&		&	111126A 	&		&		&	 x 	&		\\
			150101A 	&	150101270 	&		&		&		&	111117A 	&	 111117510 	&		&		&		\\
			141212A 	&		&	 x 	&		&		&	111020A 	&		&	 x 	&		&		\\
			141205A 	&	141205337 	&		&		&		&	110420B 	&	 110420946 	&		&		&		\\
			140930B 	&		&		&	 x 	&		&	110112A 	&		&		&		&	 x	\\
			140903A 	&		&	 x 	&		&		&	101224A 	&	 101224227 	&		&		&		\\
			140622A 	&		&	 x 	&		&		&	101219A 	&		&	 x 	&		&		\\
			140611A 	&		&		&	 x 	&		&	101129A 	&	 101129652 	&		&		&		\\
			140606A 	&		&		&		&	 x	&	100724A 	&		&	 x 	&		&		\\
			140516A 	&		&		&		&	 x	&	100702A 	&		&		&	 x 	&		\\
			140414A 	&		&		&	 x 	&		&	100628A 	&		&	 x 	&		&		\\
			140402A 	&	 140402007 	&		&		&		&	100625A 	&	 100625773 	&		&		&		\\
			140320A 	&	 140320092 	&		&		&		&	100206A 	&	 100206563 	&		&		&		\\
			140129B 	&		&		&		&	 x	&	100117A 	&	 100117879 	&		&		&		\\
			131004A 	&	 131004904 	&		&		&		&	091109B 	&		&		&	 x 	&		\\
			130912A 	&	 130912358 	&		&		&		&	090815C 	&		&		&		&	 x	\\
			130626A 	&	 130626452 	&		&		&		&	090621B 	&	 090621922 	&		&		&		\\
			130603B 	&		&	 x 	&		&		&	090515 	&		&	 x 	&		&		\\
			130515A 	&	 130515056 	&		&		&		&	090510 	&	 090510016 	&		&		&		\\
			130313A 	&		&	 x 	&		&		&	090426 	&		&	 x 	&	 x 	&		\\
			121226A 	&		&	 x 	&		&		&	090417A 	&		&	 x 	&		&		\\
			120804A 	&		&	 x 	&		&		&	090305A 	&		&		&		&	 x	\\
			120630A 	&		&	 x 	&		&		&	081226A 	&	 081226044 	&		&		&		\\
			120521A 	&		&		&	 x 	&		&	081101 	&	 081101491 	&		&		&		\\
			120403A 	&		&		&		&	 x	&	081024A 	&	 081024245 	&		&		&		\\
			120305A 	&		&	 x 	&		&		&	080919 	&		&	 x 	&		&		\\
			120229A 	&		&		&	 x 	&		&	080905A 	&	 080905499 	&		&		&		\\
			
			\hline
		\end{tabular}
	\end{adjustbox}
	\caption{BAT SGRBS as viewed by GBM. List of all BAT SGRBs within our time window with the GBM trigger number if it was also detected by the GBM. If the trigger wasn't found by GBM then it was checked if the location was occulted by the Earth, occurred during transit of the SAA, and if it was neither then the trigger was missed.}
	\label{table:BatToGBMfull}
\end{table}

\begin{table}
	\begin{adjustbox}{width=\textwidth,totalheight=\textheight,keepaspectratio}
		\begin{tabular}{| c | c | c | c | c | c | c | c | c |}
			\hline
			\multicolumn{1}{|c}{} & \multicolumn{1}{|c}{} & \multicolumn{1}{c}{} & \multicolumn{1}{c}{BAT} & \multicolumn{1}{c}{} &\multicolumn{1}{c|}{} & \multicolumn{1}{c}{} & \multicolumn{1}{c}{GBM} &\multicolumn{1}{c|}{}  \\ 
			\hline
			GRB     & $T_{90}$    & Fluence & Peak Flux & PL    & Partial    & LAT offset & \# of NaI   & Ground  \\
			(Name)  & (sec)       & (\%ile) & (\%ile)   & Index & Coding     & (degrees)  & detectors   & (signal)  \\
			\hline
			140606A & 0.34        & 34.1    &  6.7      & 0.53 &  86        &  91        & 3 (18, 49)  & Strong \\
			120403A & 1.25        & 61.0    & 40.0      & 1.64 &  50        &  71        & 3 (32, 33)  & Strong \\
			\hline
			140516A & 0.19        &  8.5    &  6.7      & 1.87 &  75        &  31        & 3 (18, 31)  & Weak   \\
			090305A & 0.40        & 52.4    & 64.0      & 0.86 &  50        &  97        & 1 (39, 69)  & Weak   \\
			\hline
			140129B & 1.36        & 47.6    & 28.0      & 2.23 & 100        &  14        & 3 (33, 53)  & None   \\
			110112A & 0.50        & 14.6    &  6.7      & 2.14 &  87        & 135        & 1 (45, 66)  & None   \\
			090815C & 0.60        & 28.0    & 16.0      & 0.90 &  76        & 116        & 1 (43, 86)  & None   \\
			\hline
		\end{tabular}
	\end{adjustbox}
	\caption{The seven BAT SGRBs untriggered in GBM. No $T_{90}$ error is greater than 0.3 s, placing them all firmly in the SGRB category. The percentile rankings are calculated from the sample of all BAT SGRBs. The average spectral power law index for a BAT SGRB is 1.23. Both measures of brightness are in the 15-150 keV band. The peak flux is the 1-second peak photon flux which is not ideal for short bursts but gives an estimate of the GRB brightness. The number of NaI detectors refers to the number of GBM NaI detectors that have the burst location within their FoV, and the numbers in parentheses are the two lowest detector angles. Strong signals are for SGRBs that GBM could claim detection alone while weak signals denote SGRBs that GBM would claim detection only if the the SGRB was also detected by another instrument.}
	
	\label{table:BatToGBMFS}
\end{table}


\end{document}